\documentclass[longauth, letter]{aa} 
\usepackage{graphicx}
\usepackage{natbib}
\usepackage{amssymb}
\usepackage{amsmath}
\usepackage{url}
\usepackage[section]{placeins}
\usepackage{multirow}
\usepackage{txfonts}
\begin{document}
   \title{Origin of the hot gas in low-mass protostars\thanks{{\it Herschel} is an ESA space observatory with science instruments provided by European-led Principal Investigator consortia and with important participation from NASA.}}
\subtitle{Herschel-PACS spectroscopy of HH 46}


\author{T.A.~van~Kempen\inst{1,2}
\and L.E.~Kristensen\inst{1}
\and G.J.~Herczeg\inst{3}
\and R.~Visser\inst{1}
\and E.F.~van~Dishoeck\inst{1,3}
\and S.F.~Wampfler\inst{4}
\and S.~Bruderer\inst{4}
\and A.O.~Benz\inst{4}
\and S.D.~Doty\inst{5}
\and C.~Brinch\inst{1}
\and M.R.~Hogerheijde\inst{1}
\and J.K.~J{\o}rgensen\inst{6}
\and M.~Tafalla\inst{7}
\and D.~Neufeld\inst{8}
\and R.~Bachiller\inst{7}
\and A.~Baudry\inst{9}
\and M.~Benedettini\inst{10}
\and E.A.~Bergin\inst{11}
\and P.~Bjerkeli\inst{12}
\and G.A.~Blake\inst{13}
\and S.~Bontemps\inst{9}
\and J.~Braine\inst{9}
\and P.~Caselli\inst{14,15}
\and J.~Cernicharo\inst{16}
\and C.~Codella\inst{15}
\and F.~Daniel\inst{16}
\and A.M.~di~Giorgio\inst{10}
\and C.~Dominik\inst{17,18}
\and P.~Encrenaz\inst{19}
\and M.~Fich\inst{20}
\and A.~Fuente\inst{21}
\and T.~Giannini\inst{22}
\and J.R.~Goicoechea\inst{16}
\and Th.~de~Graauw\inst{23}
\and F.~Helmich\inst{23}
\and F.~Herpin\inst{9}
\and T.~Jacq\inst{9}
\and D.~Johnstone\inst{24,25}
\and M.J.~Kaufman\inst{26}
\and B.~Larsson\inst{27}
\and D.~Lis\inst{28}
\and R.~Liseau\inst{12}
\and M.~Marseille\inst{23}
\and C.~M$^{\rm c}$Coey\inst{20}
\and G.~Melnick\inst{2}
\and B.~Nisini\inst{22}
\and M.~Olberg\inst{12}
\and B.~Parise\inst{29}
\and J.C.~Pearson\inst{30}
\and R.~Plume\inst{31}
\and C.~Risacher\inst{23}
\and J.~Santiago-Garc\'{i}a\inst{32}
\and P.~Saraceno\inst{10}
\and R.~Shipman\inst{23}
\and F.~van der Tak\inst{23,33}
\and F.~Wyrowski\inst{29}
\and U.A.~Y{\i}ld{\i}z\inst{1}
\and M.~Ciechanowicz\inst{29}
\and L.~Dubbeldam\inst{23}
\and S.~Glenz\inst{34}
\and R.~Huisman\inst{23}
\and R.H.~Lin\inst{29}
\and P.~Morris\inst{35}
\and J.A.~Murphy\inst{36}
\and N.~Trappe\inst{36}
}

\institute{
Leiden Observatory, Leiden University, PO Box 9513, 2300 RA Leiden, The Netherlands
\and
Harvard-Smithsonian Center for Astrophysics, 60 Garden Street, MS 42, Cambridge, MA 02138, USA
\and
Max-Planck-Institut f\"{u}r Extraterrestrische Physik, Giessenbachstrasse 1, 85748 Garching, Germany
\and
Institute of Astronomy, ETH Zurich, 8093 Zurich, Switzerland
\and
Department of Physics and Astronomy, Denison University, Granville, OH, 43023, USA
\and
Centre for Star and Planet Formation, Natural History Museum of Denmark, University of Copenhagen,
{\O}ster Voldgade 5-7, DK-1350 Copenhagen K., Denmark
\and
Observatorio Astron\'{o}mico Nacional (IGN), Calle Alfonso XII,3. 28014, Madrid, Spain
\and
Department of Physics and Astronomy, Johns Hopkins University, 3400 North Charles Street, Baltimore, MD 21218, USA
\and
Universit\'{e} de Bordeaux, Laboratoire d'Astrophysique de Bordeaux, France; CNRS/INSU, UMR 5804, Floirac, France
\and
INAF - Instituto di Fisica dello Spazio Interplanetario, Area di Ricerca di Tor Vergata, via Fosso del Cavaliere 100, 00133 Roma, Italy
\and
Department of Astronomy, University of Michigan, 500 Church Street, Ann Arbor, MI 48109-1042, USA
\and
Department of Radio and Space Science, Chalmers University of Technology, Onsala Space Observatory, 439 92 Onsala, Sweden
\and
California Institute of Technology, Division of Geological and Planetary Sciences, MS 150-21, Pasadena, CA 91125, USA
\and
School of Physics and Astronomy, University of Leeds, Leeds LS2 9JT, UK
\and
INAF - Osservatorio Astrofisico di Arcetri, Largo E. Fermi 5, 50125 Firenze, Italy
\and
Centro de Astrobiolog\'{\i}a, Departamento de Astrof\'{\i}sica, CSIC-INTA, Carretera de Ajalvir, Km 4, Torrej\'{o}n de Ardoz. 28850, Madrid, Spain
\and
Astronomical Institute Anton Pannekoek, University of Amsterdam, Kruislaan 403, 1098 SJ Amsterdam, The Netherlands
\and
Department of Astrophysics/IMAPP, Radboud University Nijmegen, P.O. Box 9010, 6500 GL Nijmegen, The Netherlands
\and
LERMA and UMR 8112 du CNRS, Observatoire de Paris, 61 Av. de l'Observatoire, 75014 Paris, France
\and
University of Waterloo, Department of Physics and Astronomy, Waterloo, Ontario, Canada
\and
Observatorio Astron\'{o}mico Nacional, Apartado 112, 28803 Alcal\'{a} de Henares, Spain
\and
INAF - Osservatorio Astronomico di Roma, 00040 Monte Porzio catone, Italy
\and
SRON Netherlands Institute for Space Research, PO Box 800, 9700 AV, Groningen, The Netherlands
\and
National Research Council Canada, Herzberg Institute of Astrophysics, 5071 West Saanich Road, Victoria, BC V9E 2E7, Canada
\and
Department of Physics and Astronomy, University of Victoria, Victoria, BC V8P 1A1, Canada
\and
Department of Physics and Astronomy, San Jose State University, One Washington Square, San Jose, CA 95192, USA
\and
Department of Astronomy, Stockholm University, AlbaNova, 106 91 Stockholm, Sweden
\and
California Institute of Technology, Cahill Center for Astronomy and Astrophysics, MS 301-17, Pasadena, CA 91125, USA
\and
Max-Planck-Institut f\"{u}r Radioastronomie, Auf dem H\"{u}gel 69, 53121 Bonn, Germany
\and
Jet Propulsion Laboratory, California Institute of Technology, Pasadena, CA 91109, USA
\and
Department of Physics and Astronomy, University of Calgary, Calgary, T2N 1N4, AB, Canada
\and
Instituto de RadioAstronom\'{i}a Milim\'{e}trica, Avenida Divina Pastora, 7, N\'{u}cleo Central E 18012 Granada, Spain
\and
Kapteyn Astronomical Institute, University of Groningen, PO Box 800, 9700 AV, Groningen, The Netherlands
\and
KOSMA, I. Physik. Institut, Universit\"{a}t zu K\"{o}ln, Z\"{u}lpicher Str. 77, D 50937 K\"{o}ln, Germany
\and
California Institute of Technology, 1200 E. California Bl., MC 100-22, Pasadena, CA. 91125  USA
\and
Experimental Physics Dept., National University of Ireland Maynooth, Co. Kildare. Ireland
}

\date{\today} \titlerunning{Herschel-PACS spectroscopy of HH~46}%


\def\placefigureSpectraNew{
\begin{figure}[!t]
\begin{center}
\includegraphics[width=\columnwidth]{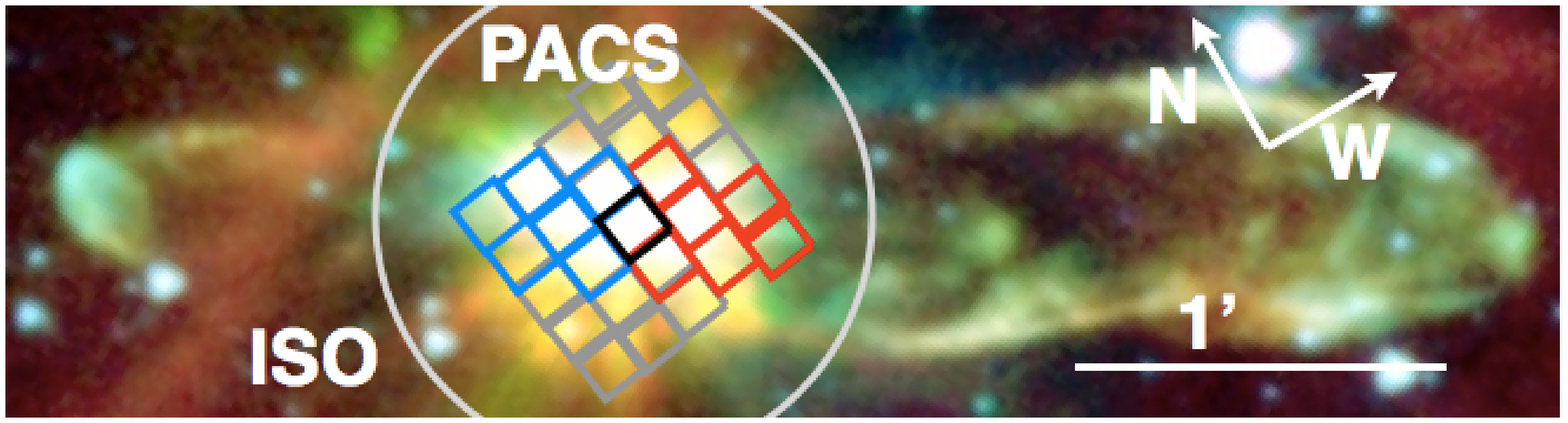}
\includegraphics[width=\columnwidth]{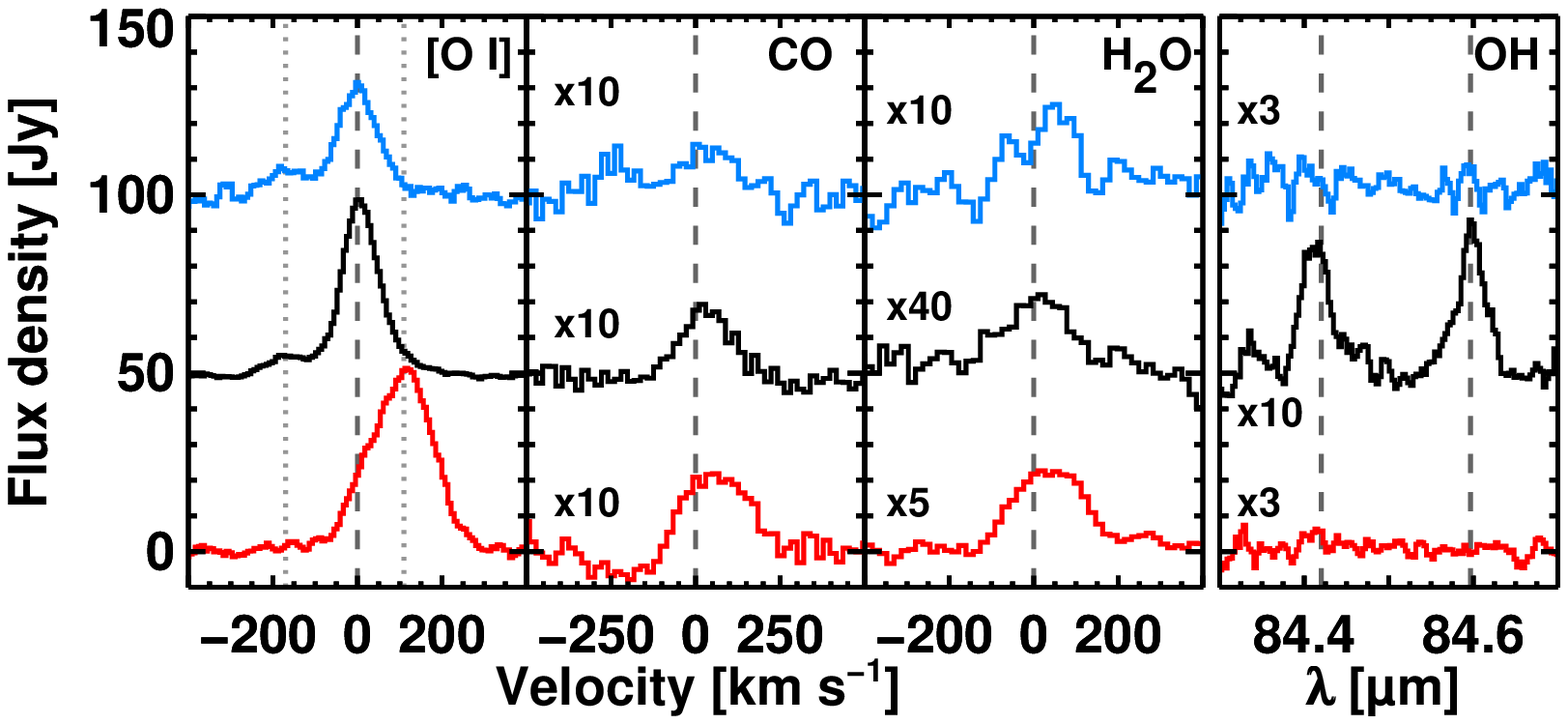}
\end{center}
\caption{The [\ion{O}{i}] 63 $\mu$m, CO 16--15 163 $\mu$m, H$_2$O $2_{12}$--$1_{01}$ 179 $\mu$m lines and the OH doublet at 84 $\mu$m (possibly blended with a small contribution from CO 31-30), at the central position (08$^{\rm h}$25$^{\rm m}$43\fs9; $-$51\degr00\arcmin36\arcsec; J2000) and integrated over the red- and blue-shifted outflow lobes. The vertical lines show the velocity of [\ion{O}{i}] emission at rest (dashed) and in the blue- and red-shifted jets (dotted). The {\it Spitzer} image of HH~46 \citep{Velusamy07} is shown with the PACS footprint and ISO beam overlaid. The spaxels covering the red- and blue-shifted lobes are indicated with coloured boxes. The central spaxel is shown in black.}
\label{fig:spec}
\end{figure}
}

\def\placeFigureCOladder{
\begin{figure}
\begin{center}
\includegraphics[width=\columnwidth]{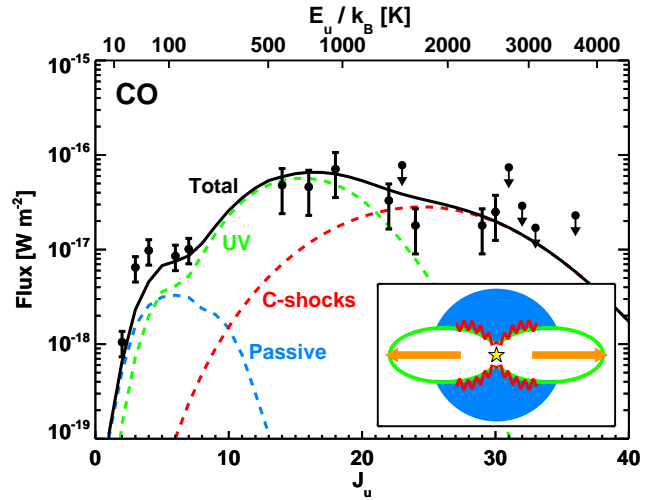}
\end{center}
\caption{CO line fluxes observed in the central PACS spaxel ($J_{\rm u}>10$) and with APEX ($J_{\rm u}<10$). Model fluxes are used to estimate the ratio of flux in a fictive PACS spaxel at the APEX wavelength and the observed APEX flux. Overplotted are predictions from a passively heated envelope (blue), a UV-heated cavity (green), and small-scale shocks in the cavity walls (red). The black line is the sum of the three. A cartoon of the different components is shown in the inset.}
\label{fig:co}
\end{figure}
}



\def\placetableone{
\begin{table}
\caption{Line fluxes observed towards HH~46 with PACS.}
\scriptsize
\centering
\begin{tabular}{l@{\extracolsep{2pt}}ccr l@{\extracolsep{2pt}}c r@{\extracolsep{0pt}}c r}
\hline \hline
Spec.       & Trans.                       & $\lambda$ ($\mu$m) & Flux$^{\rm a}$   & Spec.  & Trans.                                                                    & & $\lambda$ ($\mu$m) & Flux$^{\rm a}$ \\
\hline
CO          & 14--13                       & 186.0              & 48\phantom{$^{\rm b}$}    & H$_2$O & 2$_{21}$--2$_{12}$                                                        & & 180.5              & $<$7$^{\rm c}$ \\
            & 16--15                       & 162.8              & 46\phantom{$^{\rm b}$}    &        & 2$_{12}$--1$_{01}$                                                        & & 179.5              & 4:\phantom{$^{\rm b}$} \\
            & 18--17                       & 144.8              & 71\phantom{$^{\rm b}$}    &        & 3$_{03}$--2$_{12}$                                                        & & 174.6              & $<$22$^{\rm c}$ \\
            & 22--21                       & 118.6              & 33\phantom{$^{\rm b}$}    &        & 3$_{13}$--2$_{02}$                                                        & & 138.5              & 14\phantom{$^{\rm b}$} \\
            & 23--22                       & 113.5              & 78$^{\rm b}$              &        & 4$_{04}$--3$_{13}$                                                        & & 125.4              & $<$18$^{\rm c}$ \\
            & 24--23                       & 108.8              & 18\phantom{$^{\rm b}$}    &        & 4$_{14}$--3$_{03}$                                                        & & 113.5              & 78$^{\rm b}$ \\
            & 29--28                       & \phantom{0}90.2    & 18\phantom{$^{\rm b}$}    &        & 2$_{21}$--1$_{10}$                                                        & & 108.1              & 17\phantom{$^{\rm b}$} \\
            & 30--29                       & \phantom{0}87.2    & 25:\,                     &        & 3$_{22}$--2$_{11}$                                                        & & \phantom{0}90.0    & $<$27$^{\rm c}$ \\
            & 31--30                       & \phantom{0}84.4    & 74$^{\rm b}$              &        & 8$_{18}$--7$_{07}$                                                        & & \phantom{0}63.3    & $<$33$^{\rm c}$ \\
            & 32--31                       & \phantom{0}81.8    & $<$29$^{\rm c}$             & \multirow{2}{*}{OH$^{\rm e}$} & \multirow{2}{*}{$\frac{1}{2}$,$\frac{3}{2}$--$\frac{1}{2}$,$\frac{1}{2}$} & \multirow{2}{*}{\bigg\{} & 163.4              & $<$27$^{\rm c}$ \\
            & 33--32                       & \phantom{0}79.4    & $<$17$^{\rm c}$  &        &                                                                           & & 163.0              & 22\phantom{$^{\rm b}$} \\
            & 36--35                       & \phantom{0}72.8    & $<$23$^{\rm c}$  &        & \multirow{2}{*}{$\frac{3}{2}$,$\frac{5}{2}$--$\frac{3}{2}$,$\frac{3}{2}$} & \multirow{2}{*}{\bigg\{} & 119.4              & 44\phantom{$^{\rm b}$} \\
\ion{C}{ii} & $^2$P$_{3/2}$--$^2$P$_{1/2}$ & 157.7              & 73\phantom{$^{\rm b}$}    &        &                                                                           & & 119.2              & 38\phantom{$^{\rm b}$} \\
\ion{O}{i}  & $^3$P$_1$--$^3$P$_2$         & \phantom{0}63.2    & 1260\phantom{$^{\rm b}$}  &        & \multirow{2}{*}{$\frac{3}{2}$,$\frac{7}{2}$--$\frac{3}{2}$,$\frac{5}{2}$} & \multirow{2}{*}{\bigg\{} & \phantom{0}84.6    & 87\phantom{$^{\rm b}$} \\
            & $^3$P$_0$--$^3$P$_2$         & 145.5              & 82\phantom{$^{\rm b}$}    &        &                                                                           & & \phantom{0}84.4    & 74$^{\rm b}$ \\
Blue$^{\rm d}$ & $^3$P$_1$--$^3$P$_2$      & \phantom{0}63.2    & 162\phantom{$^{\rm b}$}   &        & \multirow{2}{*}{$\frac{1}{2}$,$\frac{1}{2}$--$\frac{3}{2}$,$\frac{3}{2}$} & \multirow{2}{*}{\bigg\{} & \phantom{0}79.2    & 38\phantom{$^{\rm b}$} \\
            &                              &                    &                  &        &                                                                           & & \phantom{0}79.1    & 55\phantom{$^{\rm b}$}\vspace{0.1em}\\
\hline
\end{tabular}
\vspace{-8pt}
\begin{list}{}{}
\item[$^{\mathrm{a}}$] Flux in the central spaxel, in 10$^{-18}$ W m$^{-2}$. Values with colons indicate weak detections or detections in only one nod. ~~~ $^{\rm b}$~ Lines blended. ~~~~~~ $^{\rm c}$ ~ 3$\sigma$ upper limit.
\item[$^{\mathrm{d}}$] Blue-shifted [\ion{O}{i}] component at $\varv=-170$ km~s$^{-1}$. ~~~~~~ $^{\rm e}$ ~ Doublet transitions.
\end{list}
\label{tab:1}
\end{table}
}

\def\placetableoneextended{
\onltab{2}{
\begin{table*}
\caption{Line fluxes observed towards HH~46 with PACS.$^{\rm a}$}
\begin{center}
\begin{tabular}{lc r@{\extracolsep{0pt}}c ccccc}
\hline \hline
& & & & & Central Spaxel & Red Outflow & Blue Outflow & All 25 Spaxels \\
\multicolumn{5}{r}{Area (square arcsec)} &  88 & 530 & 530 & 2200 \\
\hline
Species & Transition & & $\lambda_{\rm lab}$ & $E_{\rm{u}}/k_{\rm B}$ & Flux & Flux & Flux & Flux \\
        &            & & ($\mu$m)            & (K)                    & \multicolumn{4}{c}{($10^{-18}$ W m$^{-2}$)} \\
\hline
CO     & 14--13                               & &           185.999 & \phantom{0}580.5 & $\phantom{0}48\pm4$\phantom{:}   & $\phantom{0}27\pm2$\phantom{:} & $18\pm2$                       & $108\pm7$\phantom{0} \\ 
       & 16--15                               & &           162.812 & \phantom{0}751.7 & $\phantom{0}46\pm5$\phantom{:}   & $\phantom{0}27\pm2$\phantom{:} & $13\pm3$                       & $\phantom{^{\rm b}}80\pm8^{\rm b}$ \\ 
       & 18--17                               & &           144.784 & \phantom{0}945.0 & $\phantom{0}71\pm7$\phantom{:}   & $\phantom{0}27\pm2$\phantom{:} & \phantom{00}$<$14              & $\phantom{^{\rm b}0}67\pm30^{\rm b}$ \\ 
       & 22--21                               & &           118.581 &           1397.4 & $\phantom{0}33\pm7$\phantom{:}   & \phantom{00:}$<$34             & \phantom{00}$<$27              & $44\pm8$ \\ 
       & $\phantom{^{\rm c}}$23--22$^{\rm c}$ & &           113.458 &           1524.2 & $\phantom{00}78\pm11$\phantom{:} & $\phantom{0}15\pm3$\phantom{:} & \phantom{00}$<$14              & $\phantom{^{\rm b}0}78\pm13^{\rm b}$ \\ 
       & 24--23                               & &           108.763 &           1656.5 & $\phantom{0}18\pm6$\phantom{:}   & \phantom{00:}$<$14             & \phantom{00}$<$26              & $38\pm9$ \\ 
       & 29--28                               & & \phantom{0}90.163 &           2399.8 & $\phantom{0}18\pm3$\phantom{:}   & \phantom{00:}$<$26             & \phantom{00}$<$19              & $28\pm6$ \\ 
       & 30--29                               & & \phantom{0}87.190 &           2564.8 & $\phantom{0}25\pm4$:             & \phantom{00:}$<$14             & \phantom{00}$<$21              & \phantom{00}$<$55 \\
       & $\phantom{^{\rm d}}$31--30$^{\rm d}$ & & \phantom{0}84.411 &           2735.3 & $\phantom{0}74\pm6$\phantom{:}   & \phantom{00:}$<$31             & \phantom{00}$<$22              & $\phantom{^{\rm b}}55\pm7^{\rm b}$ \\ 
       & 32--31                               & & \phantom{0}81.806 &           2911.2 & \phantom{00:}$<$29               & \phantom{00:}$<$40             & \phantom{00}$<$40              & \phantom{00}$<$61 \\ 
       & 33--32                               & & \phantom{0}79.360 &           3092.5 & \phantom{00:}$<$17               & \phantom{00:}$<$27             & \phantom{00}$<$41              & \phantom{00}$<$57 \\ 
       & 36--35                               & & \phantom{0}72.843 &           3668.8 & \phantom{00:}$<$23               & \phantom{00:}$<$49             & \phantom{00}$<$51              & \phantom{0}$<$136 \\ 
\multicolumn{9}{c}{~} \\
H$_2$O     & 2$_{21}$--2$_{12}$                               & &           180.488 & \phantom{0}194.1 & \phantom{000:}$<$7               & \phantom{000:}$<$9             & \phantom{000}$<$9 & \phantom{00}$<$14 \\ 
           & 2$_{12}$--1$_{01}$                               & &           179.527 & \phantom{0}114.4 & $\phantom{00}4\pm2$:             & $\phantom{0}46\pm3$\phantom{:} & $16\pm2$          & $82\pm7$ \\ 
           & 3$_{03}$--2$_{12}$                               & &           174.626 & \phantom{0}196.8 & \phantom{00:}$<$22               & $\phantom{0}24\pm2$\phantom{:} & \phantom{00}$<$21 & $54\pm7$ \\ 
           & 3$_{13}$--2$_{02}$                               & &           138.528 & \phantom{0}204.7 & $\phantom{0}14\pm4$\phantom{:}   & $\phantom{0}18\pm3$:           & \phantom{00}$<$10 & $34\pm8$ \\ 
           & 4$_{04}$--3$_{13}$                               & &           125.354 & \phantom{0}319.5 & \phantom{00:}$<$18               & \phantom{000:}$<$6             & $14\pm3$          & \phantom{00}$<$24 \\ 
           & $\phantom{^{\rm c}}$4$_{14}$--3$_{03}$$^{\rm c}$ & &           113.537 & \phantom{0}323.5 & $\phantom{00}78\pm11$\phantom{:} & $\phantom{0}15\pm3$\phantom{:} & \phantom{00}$<$14 & $\phantom{^{\rm b}0}78\pm13^{\rm b}$ \\ 
           & 2$_{21}$--1$_{10}$                               & &           108.073 & \phantom{0}194.1 & $\phantom{0}17\pm7$\phantom{:}   & $\phantom{0}16\pm3$:           & $21\pm4$          & $\phantom{0}64\pm11$ \\ 
           & $3_{22}$--$2_{11}$                               & & \phantom{0}89.988 & \phantom{0}296.8 & \phantom{00:}$<$27               & \phantom{00:}$<$22             & \phantom{00}$<$49 & \phantom{0}$<$110 \\ 
           & 8$_{18}$--7$_{07}$                               & & \phantom{0}63.324 &           1070.7 & \phantom{00:}$<$33               & \phantom{00:}$<$10             & \phantom{00}$<$22 & \phantom{00}$<$43 \\ 
\multicolumn{9}{c}{~} \\
OH$^{\rm e}$ & \multirow{2}{*}{$\frac{1}{2}$,$\frac{3}{2}$--$\frac{1}{2}$,$\frac{1}{2}$}                               & \multirow{2}{*}{\bigg\{} &           163.396 & \phantom{0}269.8 & \phantom{00:}$<$27             & \phantom{000:}$<$4             & \phantom{000}$<$9 & \phantom{00}$<$10 \\ 
             &                                                                                                         &                          &           163.015 & \phantom{0}270.2 & $\phantom{0}22\pm4$\phantom{:} & \phantom{00:}$<$18             & \phantom{000}$<$5 & $20\pm4$ \\ 
             & \multirow{2}{*}{$\frac{3}{2}$,$\frac{5}{2}$--$\frac{3}{2}$,$\frac{3}{2}$}                               & \multirow{2}{*}{\bigg\{} &           119.441 & \phantom{0}120.5 & $\phantom{0}44\pm7$\phantom{:} & $\phantom{0}49\pm5$\phantom{:} & \phantom{000}$<$7 & $\phantom{^{\rm b}0}86\pm12^{\rm b}$ \\ 
             &                                                                                                         &                          &           119.234 & \phantom{0}120.7 & $\phantom{0}38\pm9$\phantom{:} & $\phantom{0}25\pm4$\phantom{:} & \phantom{00}$<$19 & $\phantom{0}71\pm15$ \\ 
             & \multirow{2}{*}{$\phantom{^{\rm d}}$$\frac{3}{2}$,$\frac{7}{2}$--$\frac{3}{2}$,$\frac{5}{2}$$^{\rm d}$} & \multirow{2}{*}{\bigg\{} & \phantom{0}84.597 & \phantom{0}290.5 & $\phantom{0}87\pm6$\phantom{:} & \phantom{00:}$<$14             & \phantom{00}$<$27 & $\phantom{0}95\pm10$ \\ 
             &                                                                                                         &                          & \phantom{0}84.420 & \phantom{0}291.2 & $\phantom{0}74\pm6$\phantom{:} & \phantom{00:}$<$31             & \phantom{00}$<$22 & $\phantom{^{\rm b}}55\pm7^{\rm b}$ \\ 
             & \multirow{2}{*}{$\frac{1}{2}$,$\frac{1}{2}$--$\frac{3}{2}$,$\frac{3}{2}$}                               & \multirow{2}{*}{\bigg\{} & \phantom{0}79.179 & \phantom{0}181.7 & $\phantom{0}38\pm5$\phantom{:} & \phantom{00:}$<$41             & \phantom{00}$<$37 & $\phantom{0}93\pm13$ \\ 
             &                                                                                                         &                          & \phantom{0}79.116 & \phantom{0}181.9 & $\phantom{0}55\pm7$\phantom{:} & \phantom{00:}$<$34             & \phantom{00}$<$35 & $\phantom{0}87\pm10$ \\ 
\multicolumn{9}{c}{~} \\
\ion{O}{i}  & $^3$P$_{0}$--$^3$P$_{2}$     & &           145.525 & \phantom{0}326.6 &  $\phantom{0}82\pm8$\phantom{:}  &  $\phantom{0}68\pm4$\phantom{:}  & $48\pm3$             &  $208\pm14$ \\ 
            & $^3$P$_1$--$^3$P$_2$         & & \phantom{0}63.184 & \phantom{0}227.7 &           $1260\pm54$\phantom{:} & $\phantom{0}186\pm28$\phantom{:} & $420\pm20$           &  $1870\pm80$\phantom{0} \\ 
            &                              & & \multicolumn{2}{c}{$\varv=-170$ km~s$^{-1}$ $^{\rm f}$}      & $\phantom{0}162\pm12$\phantom{:} & \phantom{0:}--                   & $90\pm8$             &  $282\pm51$ \\ 
            &                              & & \multicolumn{2}{c}{$\varv=+110$ km~s$^{-1}$ $^{\rm f}$}      & \phantom{0:}--                    & $1230\pm52$\phantom{:}           & \phantom{0}--        &  $1510\pm100$ \\ 
\multicolumn{9}{c}{~} \\
\ion{C}{ii} & $^2$P$_{3/2}$--$^2$P$_{1/2}$ & & 157.741           & \phantom{00}91.2 & $\phantom{0}73\pm6$\phantom{:}   & $\phantom{0}77\pm4$\phantom{:}   & $140\pm6$\phantom{0} &  $505\pm23$ \\ 
\hline
\end{tabular}
\end{center}
\begin{list}{}{}
\item[$^{\rm a}$] Fluxes are measured from Gaussian fits. Lines marked with a colon are only weakly detected. Listed uncertainties are 68\% confidence intervals, upper limits are 95\% confidence intervals. The uncertainties do not include the 50\% error margin in the relative spectral response function.
\item[$^{\rm b}$] The total flux is less than the combined flux from the central spaxel, the red outflow and the blue outflow. This is a result of higher noise in the spectrum summed over all spaxels, leading to a different fit and PSF over-correction.
\item[$^{\rm c}$] The CO 23--22 and H$_2$O 4$_{14}$--3$_{03}$ lines at 113.5 $\mu$m are blended.
\item[$^{\rm d}$] The CO 31--30 and OH $3/2$,$7/2$--$3/2$,$5/2$ lines at 84.4 $\mu$m are blended.
\item[$^{\rm e}$] Doublet transitions.
\item[$^{\rm f}$] Blue- and red-shifted components of the [\ion{O}{i}] 63.2 $\mu$m line. The blue component is not detected in the spaxels covering the red outflow, and vice versa. Futhermore, the red component is not detected in the central spaxel.
\label{tab:1ext}
\end{list}
\end{table*}
}
}

\def\placeTablecoolingthree{
\begin{table}
\caption{Origin of line emission and cooling rates of various species for the HH 46 central spaxel.}
\begin{center}
\begin{tabular}{l r l}
\hline \hline
Species & Cooling & Origin\\
 & 10$^{-3}$ $L_\odot$ \\ \hline
CO & \phantom{$>$}0.1 & Passively heated envelope \\
CO & \phantom{$>$}3.8 & UV-heated cavity walls \\
CO & \phantom{$>$}2.8 & $C$-type shocks \\
H$_2$O & \phantom{$>$}0.1 & Passively heated envelope \\
H$_2$O & \phantom{$>$}5.0 & UV or $C$-type shocks\\
\ion{O}{i} & \phantom{$>$}9.5 & $J$-type shock \\
OH     & $>$2.4 & $J$-type shock \\
\ion{C}{ii} & \phantom{$>$}0.1 & Surrounding cloud \\ \hline
Total & $>$23.8 \\ \hline
\end{tabular}
\end{center}
\label{tab:cool}
\end{table}
}

\abstract {} 
{`Water in Star-forming regions with \textit{Herschel}' (WISH) is a \textit{Herschel} Key Programme aimed at understanding the physical and chemical structure of young stellar objects (YSOs) with a focus on water and related species.}
{The low-mass protostar HH~46 was observed with the Photodetector Array Camera and Spectrometer (PACS) on  the \textit{Herschel} Space Observatory to measure emission in H$_2$O, CO, OH, [\ion{O}{i}], and [\ion{C}{ii}] lines located between 63 and 186 $\mu$m. The excitation and spatial distribution of emission can disentangle the different heating mechanisms of YSOs, with better spatial resolution and sensitivity than previously possible.}
{Far-IR line emission is detected at the position of the protostar and along the outflow axis. The OH emission is concentrated at the central position, CO emission is bright at the central position and along the outflow, and H$_2$O emission is concentrated in the outflow. In addition, [\ion{O}{i}] emission is seen in low-velocity gas, assumed to be related to the envelope, and is also seen shifted up to 170 km~s$^{-1}$ in both the red- and blue-shifted jets. Envelope models are constructed based on previous observational constraints. They indicate that passive heating of a spherical envelope by the protostellar luminosity cannot explain the high-excitation molecular gas detected with PACS, including CO lines with upper levels at $>$2500 K above the ground state. Instead, warm CO and H$_2$O emission is probably produced in the walls of an outflow-carved cavity in the envelope, which are heated by UV photons and non-dissociative $C$-type shocks. The bright OH and [\ion{O}{i}] emission is attributed to $J$-type shocks in dense gas close to the protostar. In the scenario described here, the combined cooling by far-IR lines within the central spatial pixel is estimated to be 2$\times$10$^{-2}$ $L_\odot$, with 60--80\% attributed to $J$- and $C$-type shocks produced by interactions between the jet and the envelope.}
{}

\keywords{Astrochemistry --- Stars: formation --- ISM: molecules --- ISM: jets and outflows --- ISM: individual objects: HH~46}

\maketitle

\section{Introduction}

The embedded phase of star formation is a critical period in the evolution of a young star, because it is the stage where the final mass of the star, the size and mass of the protoplanetary disk, and the initial chemical composition of the disk are determined \citep{Andre00,Visser09b,Jorgensen09}. Many physical processes
occur simultaneously in the immediate surroundings of the protostar: infall in the collapsing envelope, outflows sweeping up and shocking the material, and energetic (UV and X-ray) photons heating and dissociating the gas \citep{Spaans95,Bachiller99,Arce07}. Because of high extinction, these processes can only be probed at far-infrared and millimetre wavelengths, but lack of observational facilities has hampered their quantification.  The goal of the `Water In Star-forming regions with {\it Herschel}' (WISH) key programme is to use H$_2$O, CO and related species to determine the physical and chemical characteristics of young stellar objects (YSOs) as functions of evolutionary stage and across a wide range of luminosities and masses \citep[see also][]{Nisini10, Fich10}.

One of the many surprises of the Infrared Space Observatory (ISO) was the detection of highly excited CO and H$_2$O lines towards low-mass YSOs with the Long Wavelength Spectrometer \citep[LWS; e.g.,][]{Giannini99,Ceccarelli99, Nisini02}. The origin of this hot gas ($\sim$500--2000 K) has been heavily debated, with two main explanations put forward: (i) shocks extending over a large area (arcmin scale), and (ii) the envelope heated by the protostellar luminosity on scales of $2-20$\arcsec. The ISO-LWS data and subsequent modelling could not distinguish between these scenarios given the large beam of $\sim$80$''$. The smaller aperture of {\it Herschel}-PACS allows imaging of these lines at 9\farcs4, sufficient to separate the on- and off-source emission. Moreover, its higher sensitivity and higher spectral resolution allow detection of weaker lines.

This letter presents Science Demonstration Phase observations of HH~46, an isolated low-mass protostar ($L_{\rm bol}\approx16\ L_\odot$, $D\approx450$ pc) located in a dense core with a prominent outflow extending out to $\sim$2$'$. It has been imaged with the {\it Spitzer} Space Telescope \citep{Velusamy07} and ground-based sub-millimetre telescopes \citep[and references therein]{vanKempen09a}. Because of its well-defined geometry, HH~46 provides an ideal testbed for separating the different physical components contributing to the observed emission and for benchmarking new models of YSOs that can be used for other sources.

\section{Observations and results}
\label{sec:obs+res}

\placefigureSpectraNew

HH~46 was observed on 26 October 2009 with {\it Herschel} \citep{Pilbratt10} using PACS \citep{Poglitsch10} in pointed line-scan spectroscopy mode (obsid 1342186315 and 1342186316). PACS is a 5$\times$5 array of 9$\farcs$4$\times$9$\farcs$4 spatial pixels (spaxels) that cover the 53--210 $\mu$m wavelength range with a spectral resolution ranging from 1000 to 4000 (the latter only at $\sim$63 $\mu$m) in spectroscopy mode. In one exposure, a wavelength segment is observed in the first order (105--210 $\mu$m) and at the same time in the second (72--105 $\mu$m) or third order (53--72 $\mu$m). The PACS spectrum of HH~46 covers 27 segments obtained in 15 separate integrations of 350 s each. Two different nod positions, located $6^\prime$ in opposite directions from the target, were used to correct for telescopic background.

Data were reduced with HIPE v2.4.0. The relative spectral response function within each band was determined from ground calibration prior to launch. The absolute wavelength scale is accurate to 30--50 km~s$^{-1}$, depending on the position of the emission peak in the cross-dispersion axis within the slit. The absolute flux calibration below and above 100 $\mu$m was separately determined from in-flight observations of (point) calibration sources. The uncertainty in absolute and relative fluxes is estimated to be 50\%, based on a comparison with the ISO-LWS continuum data from \citet{Nisini02}.

\placetableone 

\placetableoneextended

Emission is detected in lines of H$_2$O, CO, OH, \ion{O}{i}, and \ion{C}{ii} (Table \ref{tab:1}, Fig.\ \ref{fig:spec}). \citet{Nisini02} detected only the [\ion{O}{i}] 63 and 145 $\mu$m and [\ion{C}{ii}] 158 $\mu$m lines with ISO. The typical $3\sigma$ sensitivity is $\sim$10$^{-17}$ W m$^{-2}$ to an unresolved emission line and unlike the [\ion{O}{i}] line, the molecular lines are all unresolved. The PACS emission in most lines and in the continuum is strongly peaked at the central position. Emission in many lines is also seen along the red and blue outflow lobes including bright emission in the spaxel centred 11$^{\prime\prime}$ SW of the source. Tables \ref{tab:1} and \ref{tab:1ext} (the latter available online) list the line fluxes at the source position, in the red and blue outflows, and in the total field-of-view. The fluxes in the central spaxel were corrected for the point-source PSF. Fluxes for the outflows were measured over the spaxels indicated in Fig.\ \ref{fig:spec} and corrected for the leaking of light from the central spaxel into adjacent spaxels.

Nine lines of water are detected in the central spaxel with $E_{\rm{u}}/k_{\rm B}$ ranging from 114 to 320 K. Most H$_2$O emission lines peak at the location of strong outflow emission 11$^{\prime\prime}$ SW (Table \ref{tab:1ext}). Emission in CO lines ranging from $J_{\rm u}=14$ to 30 ($E_{\rm u}/k_{\rm B}=580$--2600 K) is seen in both the central spaxel and in the outflows (Tables \ref{tab:1} and \ref{tab:1ext}; Fig.\ \ref{fig:co}). Strong OH emission is detected in the four doublets at 79, 84, 119, and 163 $\mu$m, arising from levels up to 290 K above the ground state. The OH emission is strongly concentrated on-source, although emission in the 119 $\mu$m doublet is also seen in the direction of the red outflow.

Emission in the [\ion{O}{i}] 63 $\mu$m and [\ion{C}{ii}] 158 $\mu$m lines is found to be extended over most of the PACS field-of-view. The total flux in these two lines is about 1.5 and 7 times weaker, respectively, in the PACS field ($50''$$\times$$50''$) than it was in the ISO-LWS beam ($\sim$80$''$). Thus, most of the [\ion{C}{ii}] and some of the [\ion{O}{i}] emission must be located within the ISO field but beyond that of PACS or in a relatively smooth background extending out to the PACS nod positions.

The high PACS spectral resolution at 63 $\mu$m of $\sim$100 km s$^{-1}$ allows the [\ion{O}{i}] 63 $\mu$m emission to be resolved into three velocity components (see Fig.\ \ref{fig:spec}). In addition to the main component around $\varv\approx0$ km~s$^{-1}$ peaking on-source, strong red- and blue-shifted [\ion{O}{i}] 63 $\mu$m emission is detected at $110$ and $-$170 km~s$^{-1}$, respectively, in several spaxels to the SW and NE of the central source. Velocities are consistent with the jet velocities measured in near-IR and optical lines \citep[e.g.,][]{Nishikawa08,Garcia10}.

\section{Analysis}
\label{sec:analysis}

Previous observations of HH~46 have revealed many properties of the circumstellar environment including the dense centrally concentrated envelope containing the protostar HH 46 IRS, the presence of warm gas along the outflow walls heated by UV radiation \citep{vanKempen09a}, the shape and size of the outflow cavities, and the presence of jets and shocks \citep[e.g.,][]{Velusamy07}.  The temperature (10--250 K) and density ($\sim$10$^4$--10$^9$ cm$^{-3}$) structure of the passively heated envelope have been determined in spherical symmetry by fitting dust radiative transfer models to the spectral enenergy distribution of the source and the spatial extent of the continuum emission. Using these properties, a set of existing models \citep{Kristensen08,vanKempen09a,Bruderer09} is adapted to predict the emission in lines of CO, H$_2$O, OH, and \ion{O}{i}. In the following, focus is placed on analysing emission from the central spaxel.

Line emission is expected to originate in the known circumstellar components: the passively heated spherical envelope, the UV-heated cavity walls, the small-scale shocks along the cavity walls, the jet, and the disk.  The jet-driven $J$-type shock is discussed in Sects. 3.3 and 3.4. Based on the PACS observations of the HD~100546 disk \citep{Sturm10}, any disk contribution is likely negligible at the distance of HH~46; hence, only the first three components are expected to cause the molecular emission.

For the spherical envelope, the model of \citet{vanKempen09a} is rerun with the new 3D non-LTE radiative transfer code LIME (Brinch \& Hogerheijde in prep.) to obtain fluxes of the higher-$J$ CO lines. The second component, the UV-heated gas in the outflow cavity walls \citep{Spaans95,vanKempen09a}, is modelled according to the method of \citet{Bruderer09}. The basis is the same spherical envelope profile, but a 65\,000 AU $\times$ 13\,000 AU ellipsoidal cavity is now carved out at an inclination of 53\degr{} \citep{Velusamy07, Nishikawa08}. The only free parameter in this scenario is the protostellar FUV luminosity, which is assumed to be 0.1 $L_\odot$ (i.e., $G_0\approx 10^4$ at 100 AU with respect to the interstellar radiation field). The gas temperature in the cavity walls is parameterised from the grid of PDR models by \cite{Kaufman99} and is typically a few hundred K; the dust temperature and density profiles are unchanged from the spherical model. More details will be reported by Visser et al. (in prep.), who will explore a wider parameter space to assess the viability of other scenarios.

Small-scale shocks created by the outflow along the cavity walls are the third  component considered for the molecular emission. Their temperature is typically a few thousand K. The shock emission is computed by tiling a number of 1D $C$-type shock models along the 2D cavity shape \citep{Kristensen08}, taking the width of each shock to be the region over which the considered species contributes significantly to the cooling in 1D shock models.  This can effectively be considered as an estimate of the beam filling factor.  For each density in the range 10$^4$--10$^{6.5}$ cm$^{-3}$, the emission is computed using the results from \citet{Kaufman96}. The only free parameter in this model is the shock velocity, which is assumed constant along the walls. For the case of CO, a velocity of 20 km s$^{-1}$ reproduces the observations, and this velocity is adopted for the other species as well.

\subsection{CO}

The PACS data (Sect. \ref{sec:obs+res}) are complemented by spectrally resolved $J_{\rm u}\leq7$ lines ($E_{\rm u}/k_{\rm B}\leq155$ K) detected with APEX \citep[$\Delta\varv\approx10$ km s$^{-1}$;][]{vanKempen09a}. The CO/H$_2$ abundance ratio in the model is taken to be 1.6$\times$10$^{-4}$ above 20 K and below $10^5$ cm$^{-3}$. In colder regions, freeze-out lowers the CO abundance by a factor of 100. 

The model spectra are convolved with the PACS beam at the relevant wavelength and extracted from a $9\farcs4$ square spaxel at the centre. Figure \ref{fig:co} shows the observations, together with the model predictions from the passive envelope, the UV-heated cavity walls, and the small-scale $C$-type shocks. Individually, each component only fits part of the data, but together they reproduce the observations over the entire range of rotational levels from $J_{\rm u}=2$--32. The results confirm the plausibility of the scenario without excluding other solutions not investigated here.

\subsection[H$_2$O]{H$_\mathit{2}$O}

The passive envelope underproduces the observed H$_2$O fluxes by two orders of magnitude. Predicting fluxes from any 2D model such as the UV-heated cavity walls is uncertain by an order of magnitude due to challenges of radiative transfer modelling of H$_2$O. Within these uncertainties, both the UV-heated cavity model and the $C$-type shock model are able to reproduce the observations independently. The former requires an H$_2$O gas abundance of only $\sim$10$^{-7}$ in the cavity walls and $\sim$10$^{-8}$ in the rest of the envelope, as expected from chemical models including photodissociation and freeze-out. The $C$-type shock component matches the observations if the abundances from \citet{Kaufman96} are scaled down to $\sim$ 7$\times$10$^{-5}$. This could be accomplished by photodissociation of H$_2$O in the shocked gas. More detailed modelling, including spatially extended emission and a comparison with spectrally resolved line profiles observed with HIFI, is needed to distinguish these scenarios.

\placeFigureCOladder

\subsection[\[O \textsc{i}\]]{[O \i]}

Within the passively heated envelope and cavity walls, the 2D models presented above yield an [\ion{O}{i}] 63 $\mu$m emission line that is narrow (3--4 km~s$^{-1}$), optically thick, and weaker than the observed flux by more than three orders of magnitude, for an \ion{O}{i} abundance of $10^{-6}$. $C$-type shocks fail to reproduce the observations by seven orders of magnitude, and are still insufficient even if all H$_2$O is dissociated into \ion{O}{i}. The observed emission is, however, reproduced in a dissociative $J$-type shock  \citep{Neufeld89b}. Thus, for the central spaxel, where bright emission is seen at the systemic velocity, [\ion{O}{i}] may trace the impact of the high-velocity jet on the densest parts of the envelope ($n_{\rm H} > 10^7$ cm$^{-3}$), causing rapid deceleration from $\varv \geq$ 200 km~s$^{-1}$ while cooling the gas through [\ion{O}{i}] emission. The observed [\ion{O}{i}] 63/145 $\mu$m line ratio is $\sim$16 in the central spaxel. Shock models predict ratios of 14--20, where the lower ratio applies to higher densities \citep[10$^6$ cm$^{-3}$;][]{Neufeld89b}. The high-$\varv$ [\ion{O}{i}] emission observed in both the central and adjacent spaxels can be produced in fast, dissociative shocks in the much lower density jet itself.

\subsection{OH}

The OH line ratios were modelled separately for a single-component slab model using an escape probability code with absorbing and emitting dust continuum (similar to Bruderer et al.\ subm.). Comparison with the observed line ratios and intensities shows that the OH lines likely originate in a high-density and high-temperature region ($n_{\rm H}$$\sim$10$^{7}$ cm$^{-3}$, $T_{\rm gas}$$>$800 K). Along with the lack of extended emission, this rules out an origin in the photon-heated cavity walls and $C$-type shocks along the cavity walls. An OH column density of $\sim$10$^{16}$ cm$^{-2}$ gives the best fit to the observations along with a physical size of the OH emitting region of 0\farcs5 ($\sim$250 AU). Other solutions are possible and will be discussed in a forthcoming paper (Wampfler et al. in prep.). 

The models indicate that better fits are obtained for temperatures higher than what can be accounted for in a passive envelope model. The major competing coolant to \ion{O}{i} in a fast, dissociative shock is OH \citep{Neufeld89b}, and it is therefore likely that some of the OH emission is caused by the jet shock impinging on the inner, dense envelope. Further modelling, including combinations of $C$- and $J$-type shocks \citep{Snell05}, will be explored to constrain the origin of the OH emission.

\subsection{Origin of hot gas}

Table \ref{tab:cool} summarizes the assigment of the various species to the different proposed physical components. It also includes the total far-IR cooling through H$_2$O, CO, OH, and [\ion{O}{i}] lines in the central spaxel. Within our scenario, cooling by CO takes place almost equally through the UV-heated cavity walls and the small-scale $C$-type shocks. The observed H$_2$O emission can be accounted for either by UV heating of the cavity walls or by $C$-type shocks, or a combination. It is impossible to distinguish between the two scenarios at present; however, the total cooling is 5$\times$10$^{-3}\ L_\odot$ in both cases using the above models to account for the non-observed line emission.
  
Cooling through [\ion{O}{i}] emission takes place on very small spatial scales, probably related directly to the jet impinging on the envelope walls. OH emission likely arises in the same dissociative shock. The OH cooling in observed lines is 2.4$\times 10^{-3}$ $L_{\odot}$, but the total OH cooling can be an order of magnitude higher if an excitation temperature of 200 K is used to account for non-observed lines. Thus, the total cooling caused by $J$- and $C$-type shocks is at least 1.5$\times$10$^{-2}$~$L_\odot$ or 60\% of the total far-IR line cooling.  The two other components (passive and UV) are responsible for the remaining 40\% of the total cooling. ISO-LWS found typical far-IR line cooling rates of 1--5$\times$10$^{-2}~L_\odot$ \citep{Nisini02}, comparable to what is observed here. The total cooling of the entire HH~46 system is higher, since only the central spaxel is considered here.

\placeTablecoolingthree

In summary, the {\it Herschel}-PACS data allow for disentangling and quantifying the energetic processes occurring in deeply embedded protostars. The models indicate, that of the two scenarios previously proposed based on ISO-LWS data, shocks are more important than the passively heated envelope in powering the far-infrared lines. Another component, the UV-heated cavity walls, also plays an important role in producing line emission. HIFI observations resolving emission lines are planned to test the proposed scenario in the framework of the WISH key programme.

\begin{acknowledgements}
This work is made possible thanks to the HIFI guaranteed time programme and the PACS instrument builders. We thank Javier Gracia Carpio, Jeroen Bouwman, Bruno Mer{\'{\i}}n, and Bart VandenBussche for help with the PACS data reduction and many funding agencies for financial support. We also thank the referee for comments that improved this paper.

\end{acknowledgements}
\small
\bibliographystyle{aa}
\bibliography{biblio}

\begin{thebibliography}{24}
\expandafter\ifx\csname natexlab\endcsname\relax\def\natexlab#1{#1}\fi

\bibitem[{{Andr\'e} {et~al.}(2000){Andr\'e}, {Ward-Thompson}, \&
  {Barsony}}]{Andre00}
{Andr\'e}, P., {Ward-Thompson}, D., \& {Barsony}, M. 2000, in Protostars \&
  Planets IV, ed. V.~{Mannings}, A.~P. {Boss}, \& S.~S. {Russell} (Tucson:
  UAP), 59

\bibitem[{{Arce} {et~al.}(2007){Arce}, {Shepherd}, {Gueth}, {Lee}, {Bachiller},
  {Rosen}, \& {Beuther}}]{Arce07}
{Arce}, H.~G., {Shepherd}, D., {Gueth}, F., {et~al.} 2007, in Protostars \&
  Planets V, ed. B.~{Reipurth}, D.~{Jewitt}, \& K.~{Keil} (Tucson: UAP), 245

\bibitem[{{Bachiller} \& {Tafalla}(1999)}]{Bachiller99}
{Bachiller}, R. \& {Tafalla}, M. 1999, in The Origin of Stars and Planetary
  Systems, ed. C.~J. {Lada}, , \& N.~D. {Kylafis} (Kluwer, Dordrecht), 227

\bibitem[{{Bruderer} {et~al.}(2009){Bruderer}, {Benz}, {Doty}, {van Dishoeck},
  \& {Bourke}}]{Bruderer09}
{Bruderer}, S., {Benz}, A.~O., {Doty}, S.~D., {van Dishoeck}, E.~F., \&
  {Bourke}, T.~L. 2009, \apj, 700, 872

\bibitem[{{Ceccarelli} {et~al.}(1999){Ceccarelli}, {Caux}, {Loinard},
  {Castets}, {Tielens}, {Molinari}, {Liseau}, {Saraceno}, {Smith}, \&
  {White}}]{Ceccarelli99}
{Ceccarelli}, C., {Caux}, E., {Loinard}, L., {et~al.} 1999, \aap, 342, L21

\bibitem[{{Fich et al.}(2010)}]{Fich10}
{Fich et al.} 2010, \aap, this volume

\bibitem[{{Garcia Lopez} {et~al.}(2010){Garcia Lopez}, {Nisini},
  {Eisl{\"o}ffel}, {Giannini}, {Bacciotti}, \& {Podio}}]{Garcia10}
{Garcia Lopez}, R., {Nisini}, B., {Eisl{\"o}ffel}, J., {et~al.} 2010, \aap,
  511, A5

\bibitem[{{Giannini} {et~al.}(1999){Giannini}, {Lorenzetti}, {Tommasi},
  {Nisini}, {Benedettini}, {Pezzuto}, {Strafella}, {Barlow}, {Clegg}, {Cohen},
  {di Giorgio}, {Liseau}, {Molinari}, {Palla}, {Saraceno}, {Smith},
  {Spinoglio}, \& {White}}]{Giannini99}
{Giannini}, T., {Lorenzetti}, D., {Tommasi}, E., {et~al.} 1999, \aap, 346, 617

\bibitem[{{J{\o}rgensen} {et~al.}(2009){J{\o}rgensen}, {van Dishoeck},
  {Visser}, {Bourke}, {Wilner}, {Lommen}, {Hogerheijde}, \&
  {Myers}}]{Jorgensen09}
{J{\o}rgensen}, J.~K., {van Dishoeck}, E.~F., {Visser}, R., {et~al.} 2009,
  \aap, 507, 861

\bibitem[{{Kaufman} \& {Neufeld}(1996)}]{Kaufman96}
{Kaufman}, M.~J. \& {Neufeld}, D.~A. 1996, \apj, 456, 611

\bibitem[{{Kaufman} {et~al.}(1999){Kaufman}, {Wolfire}, {Hollenbach}, \&
  {Luhman}}]{Kaufman99}
{Kaufman}, M.~J., {Wolfire}, M.~G., {Hollenbach}, D.~J., \& {Luhman}, M.~L.
  1999, \apj, 527, 795

\bibitem[{{Kristensen} {et~al.}(2008){Kristensen}, {Ravkilde}, {Pineau des
  For{\^e}ts}, {Cabrit}, {Field}, {Gustafsson}, {Diana}, \&
  {Lemaire}}]{Kristensen08}
{Kristensen}, L.~E., {Ravkilde}, T.~L., {Pineau des For{\^e}ts}, G., {et~al.}
  2008, \aap, 477, 203

\bibitem[{{Neufeld} \& {Dalgarno}(1989)}]{Neufeld89b}
{Neufeld}, D.~A. \& {Dalgarno}, A. 1989, \apj, 344, 251

\bibitem[{{Nishikawa} {et~al.}(2008){Nishikawa}, {Takami}, {Hayashi},
  {Wiseman}, \& {Pyo}}]{Nishikawa08}
{Nishikawa}, T., {Takami}, M., {Hayashi}, M., {Wiseman}, J., \& {Pyo}, T. 2008,
  \apj, 684, 1260

\bibitem[{{Nisini} {et~al.}(2002){Nisini}, {Giannini}, \&
  {Lorenzetti}}]{Nisini02}
{Nisini}, B., {Giannini}, T., \& {Lorenzetti}, D. 2002, \apj, 574, 246

\bibitem[{{Nisini et al.}(2010)}]{Nisini10}
{Nisini et al.} 2010, \aap, this volume

\bibitem[{{Pilbratt et al.}(2010)}]{Pilbratt10}
{Pilbratt et al.} 2010, \aap, this volume

\bibitem[{{Poglitsch et al.}(2010)}]{Poglitsch10}
{Poglitsch et al.} 2010, \aap, this volume

\bibitem[{{Snell} {et~al.}(2005){Snell}, {Hollenbach}, {Howe}, {Neufeld},
  {Kaufman}, {Melnick}, {Bergin}, \& {Wang}}]{Snell05}
{Snell}, R.~L., {Hollenbach}, D., {Howe}, J.~E., {et~al.} 2005, \apj, 620, 758

\bibitem[{{Spaans} {et~al.}(1995){Spaans}, {Hogerheijde}, {Mundy}, \& {van
  Dishoeck}}]{Spaans95}
{Spaans}, M., {Hogerheijde}, M.~R., {Mundy}, L.~G., \& {van Dishoeck}, E.~F.
  1995, \apjl, 455, 167

\bibitem[{{Sturm et al.}(2010)}]{Sturm10}
{Sturm et al.} 2010, \aap, this volume

\bibitem[{{van Kempen} {et~al.}(2009){van Kempen}, {van Dishoeck},
  {G{\"u}sten}, {Kristensen}, {Schilke}, {Hogerheijde}, {Boland}, {Nefs},
  {Menten}, {Baryshev}, \& {Wyrowski}}]{vanKempen09a}
{van Kempen}, T.~A., {van Dishoeck}, E.~F., {G{\"u}sten}, R., {et~al.} 2009,
  \aap, 501, 633

\bibitem[{{Velusamy} {et~al.}(2007){Velusamy}, {Langer}, \&
  {Marsh}}]{Velusamy07}
{Velusamy}, T., {Langer}, W.~D., \& {Marsh}, K.~A. 2007, \apjl, 668, L159

\bibitem[{{Visser} {et~al.}(2009){Visser}, {van Dishoeck}, {Doty}, \&
  {Dullemond}}]{Visser09b}
{Visser}, R., {van Dishoeck}, E.~F., {Doty}, S.~D., \& {Dullemond}, C.~P. 2009,
  \aap, 495, 881

\end{thebibliography}

\end{document}